\documentclass[a4paper,11pt]{article}
\usepackage{jcappub}
\usepackage{lineno}

\usepackage[T1]{fontenc} 
\usepackage{caption}
\usepackage{subcaption}
\usepackage{graphicx,verbatim}
\usepackage{mathrsfs}
\usepackage{amsmath}
\usepackage{amssymb}
\usepackage{mathtools} 
\usepackage{soul}
\usepackage{graphicx}
\usepackage{dcolumn}
\usepackage{bm}
\usepackage[colorlinks=true,linkcolor=blue,citecolor=blue,urlcolor=blue,linktocpage=true]{hyperref}
\usepackage{cases}

\title{\boldmath The window on heavy charged dark matter was never open}

\author[a]{Daniele Perri,}
\author[b]{Glennys Farrar}

\affiliation[a]{Institute of Theoretical Physics, Faculty of Physics, University of
Warsaw, \\ ul. Pasteura 5, PL-02-093 Warsaw, Poland\\}

\affiliation[b]{Center for Cosmology and Particle Physics, Department of Physics, New York University, \\ New York, NY 10003, USA}

\emailAdd{daniele.perri@fuw.edu.pl}

\abstract{There is a claim in the literature that charged dark matter particles in the mass range $100 (q_{\rm X}/e)^2~\mathrm{TeV} \leq m_{\rm X} \leq 10^8 (q_{\rm X}/e)~\mathrm{TeV}$ are allowed,
based on arguing that heavy charged particles cannot reach the Earth from outside the magnetized region of the Milky Way (Chuzhoy-Kolb, 2009).  We point out that this claim fails for physical models for the Galactic magnetic field. 
We explicitly confirm our argument by simulating with the software CRPropa the trajectories of heavy charged dark matter in models of the Galactic magnetic field.
}

\begin{document}
\maketitle
\flushbottom

\section{Introduction}
\label{sec:intro}

The quest for good candidates for dark matter has decades of history, with varying success. The possibility of charged massive particles (CHAMPs) acting as a component of the dark matter is strongly constrained by the limits on dark matter self-interaction (see for example \cite{PhysRevD.88.117701}). However, stable charged particles can avoid the constraints for interacting dark matter if the particle mass is sufficiently large and the required number density consequently small enough \cite{DERUJULA1990173, PICHARD1987383, Kobayashi:2023ryr}. 
Additionally, CHAMP dark matter is also constrained by the results of terrestrial and satellite cosmic-ray detectors \cite{Taoso_2008, PhysRevD.83.063509, Dunsky_2019, PhysRevLett.120.211804}, which exclude the possibility of large CHAMP flux within the Milky Way (see \cite{Cirelli:2024ssz} for a recent review).

In \cite{Chuzhoy:2008zy}, the authors claimed that these direct detection constraints on charged dark matter do not apply to particles with mass in the range $100 (q_{\rm X}/e)^2~\mathrm{TeV} \leq m_{\rm X} \leq 10^8 (q_{\rm X}/e)~\mathrm{TeV}$ as a consequence of a local depletion of the dark matter abundance within the magnetic region of the Milky Way. 
The authors argued that such a depletion naturally occurs due to the interaction of the CHAMPs with the Galactic magnetic fields. 
Specifically, they assert that charged particles in this mass range would be unable to reach the Earth from the Galactic halo (outside the magnetized disk).
Moreover, interactions between CHAMPs and expanding supernova remnants within the Galactic disk would boost the kinetic energy of CHAMPs inside the Galactic disk and expel them on a time-scale of $0.1 - 1$ Gyrs.

In this paper we demonstrate how the argument of \cite{Chuzhoy:2008zy} that CHAMPs cannot enter the Galactic disk depends on the highly nonphysical Galactic magnetic field model considered by those authors. 
We simulate the evolution of CHAMPs outside the Galactic disk for several different models of the Galactic magnetic field (GMF), chosen to make different points: a toroidal field obtained from a toroidal solenoid crossed by an electric current, the data-driven JF12 model \cite{Jansson_2012, Kleimann_2019}, and a finite-size homogeneous magnetic field. The numerical simulations confirm our thesis, that the conclusion of~\cite{Chuzhoy:2008zy} results from their choice of a grossly unphysical model of the GMF. 

There is a sentiment in the community that Liouville's theorem implies that a uniform, isotropic cosmic ray distribution outside the Galaxy necessarily implies an isotropic distribution at an arbitrary position within the Galaxy if energy losses during propagation are negligible, e.g.,~\cite{Farrar:2014hma, deOliveira:2023kvu}.  This cannot be strictly true, as illustrated by the following example.  Suppose that the Galaxy is surrounded by a sphere which perfectly reflects cosmic rays.  That setup conserves energy and satisfies Liouville's theorem, since the phase space density is conserved along all trajectories.  However, this is a special degenerate example since if there were even a tiny hole in the sphere, CRs would pass through it and the sphere would become uniformly filled.  

This paper is organized as follows. In Section \ref{sec:galaxy} we introduce the models of the Galactic magnetic field considered in our discussion and we discuss the argument of \cite{Chuzhoy:2008zy} based on the gyroradius of the CHAMPs. 
In Section \ref{sec:simulation} we discuss the results of the simulations of the trajectories of CHAMPs in the Galactic magnetic field. We then conclude in Section \ref{sec:conclusion}.
In Appendix~\ref{sec:number_estimate}, we discuss the characteristic of Lambert's distribution used for the particle velocity distribution in our simulations.
Throughout this paper, we use Heaviside-Lorentz units, with $c = \hbar = k_B = 1$, and use $M_{\rm Pl}$ to denote the reduced Planck mass~$ (8 \pi G)^{-1/2}$.

\section{Charged dark matter in the Galactic disk}
\label{sec:galaxy}

The motion of CHAMPs moving inside the Galactic disk is affected by the presence of the Galactic magnetic field, which has a typical strength of $B_{\rm G} \approx 2 \times 10^{-6}~\mathrm{G}$.
The equation of motion of CHAMPs with mass $m_{\rm X}$ and charge $q_{\rm X} = e$, moving at velocity $\mathbf{v}$ in the Galactic magnetic field is given by the Lorentz force,
\begin{equation}
    \frac{d}{dt} \left(\gamma m_{\rm X} \mathbf{v} \right) = q_{\rm X} \mathbf{v} \times \mathbf{B}_{\rm G} .
\end{equation}
In this section, we describe the models of the Galactic magnetic field, on which we base the numerical simulations of the next sections, and point out the characteristic gyroradius of CHAMPs in the Milky Way.

\subsection{Models of the Galactic Magnetic Field}
\label{sec:models}

\begin{figure}
    \centering
    \begin{subfigure}[b]{0.49\textwidth}
         \centering
         \includegraphics[width=\textwidth]{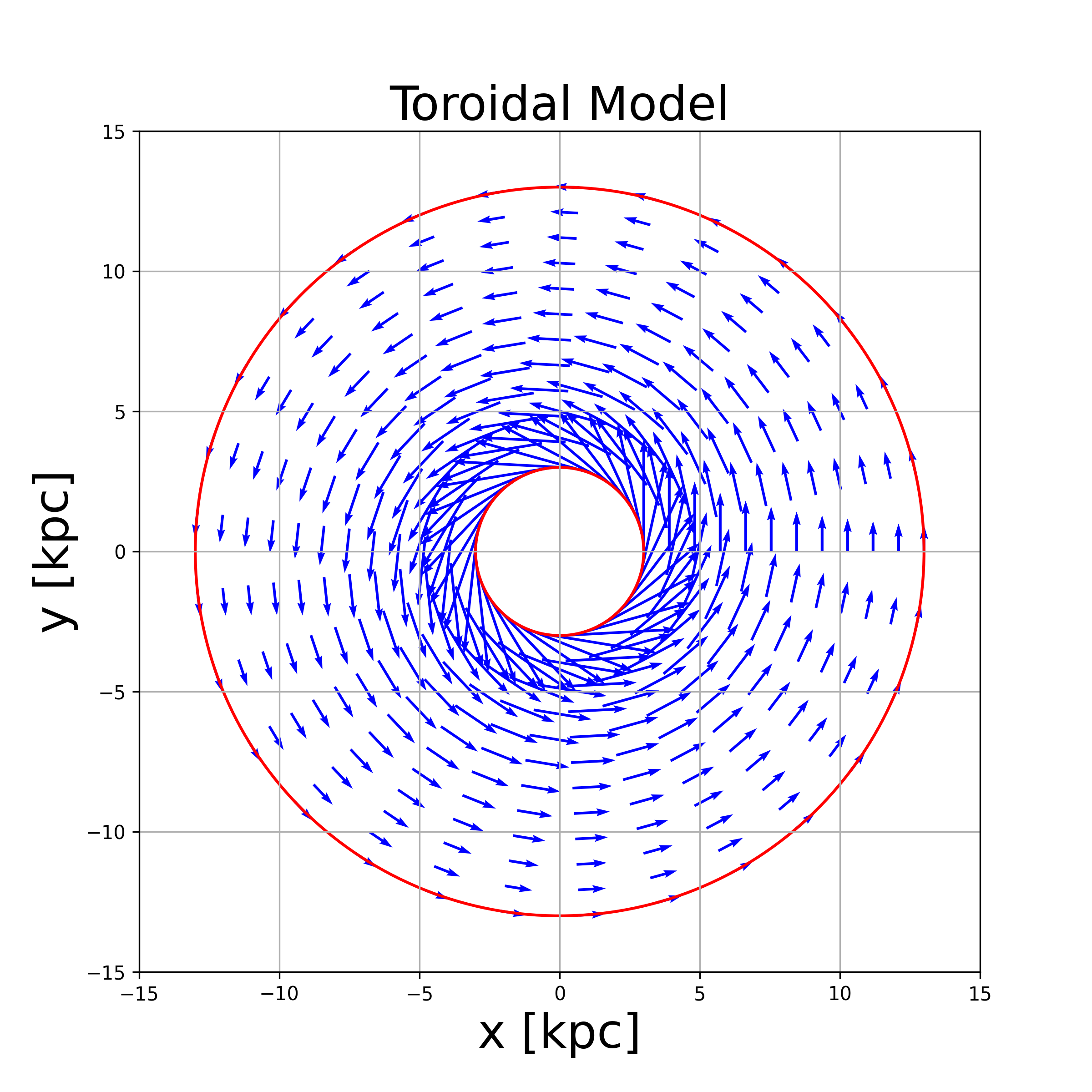}
         \caption{}
         \label{fig:toroidal}
     \end{subfigure}
     \hfill
     \begin{subfigure}[b]{0.49\textwidth}
         \centering
         \includegraphics[width=\textwidth]{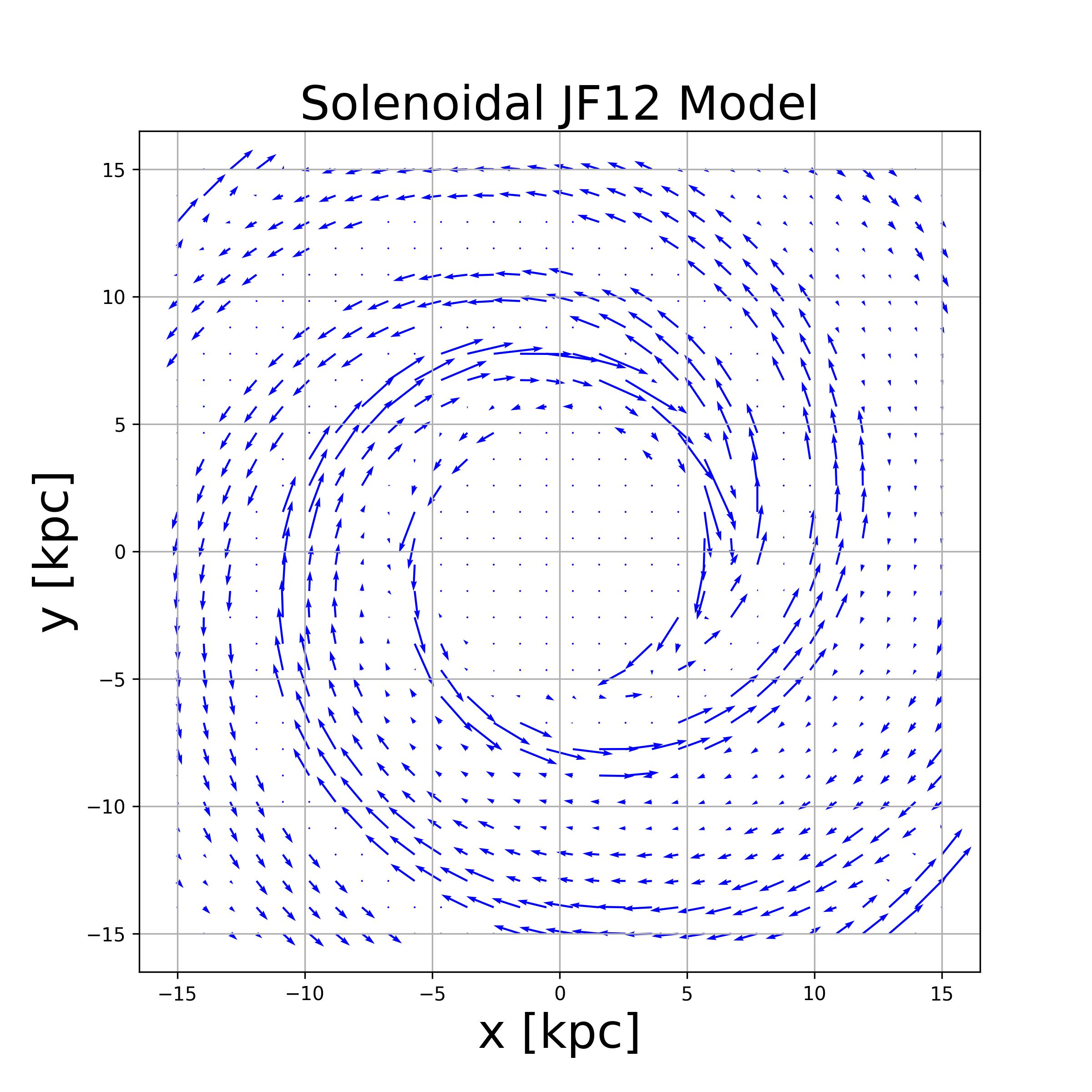}
         \caption{}
         \label{fig:JF12}
     \end{subfigure}
     \hfill
     \begin{subfigure}[b]{0.49\textwidth}
         \centering
         \includegraphics[width=\textwidth]{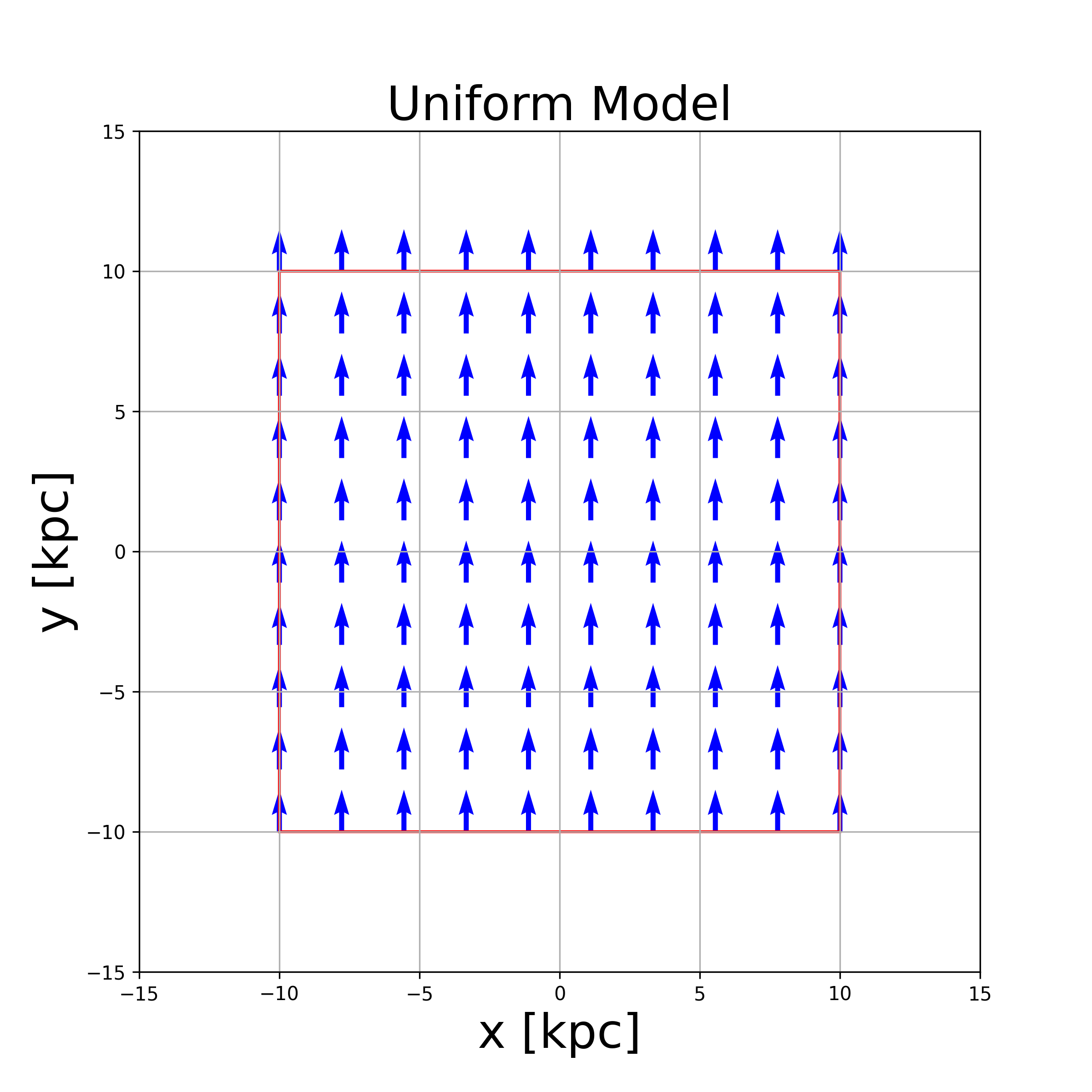}
         \caption{}
         \label{fig:uniform}
     \end{subfigure}
     \caption{Models of the Galactic magnetic field in the $xy$-plane. (a): toroidal magnetic field as in Eq.~\eqref{eq:toroidal} with $R_1 = 5~\mathrm{kpc}$ and $R_2 = 8~\mathrm{kpc}$; (b): disk component of the solenoidal JF12 model as in \cite{Kleimann_2019}; (c): finite version of the uniform magnetic field model adopted in \cite{Chuzhoy:2008zy} as in Eq.~\eqref{eq:uniform} with $R = 5~\mathrm{kpc}$ and $R' = 10~\mathrm{kpc}$.}
     \label{fig:fields}
\end{figure}

The models of the Galactic magnetic field discussed in this work are shown in Figure~\ref{fig:fields}. We now discuss each of them in detail.

The first model corresponds to the magnetic field produced by a toroidal solenoid with inner radius $R_1$ and central radius $R_2$ crossed by an electric current (Figure~\ref{fig:toroidal}). Such a system presents an analytic solution for the magnetic field. Here we show the expression for the solution in cylindrical coordinates:
\begin{equation}
\label{eq:toroidal}
\mathbf{B}_{\rm G}(r) =
  \begin{dcases}
B_{\rm toro} \frac{R_2}{r} \mathbf{\hat{\theta}}
  & \mathrm{if}\, \, \,  
\left(r - R_2\right)^2 + z^2 \leq R_1^2,
 \\
0
  & \mathrm{else}\, \, \, .
 \end{dcases}
\end{equation}
The field amplitude is therefore non-zero only inside the solenoid and depends only on the distance from the central axis of the torus.

The second model is the JF12 model \cite{Jansson_2012} in the solenoidal improved version presented in \cite{Kleimann_2019} (Figure~\ref{fig:JF12}), which represents a good phenomenological model of the Galactic magnetic field within the Galactic disk.
The model consists of a large-scale regular field (with three different components: disk, toroidal halo and poloidal halo fields), a large-scale random striated field, and a small-scale random turbulent field.
The field is described using the standard galactocentric coordinate system, where the Galactic center is at the origin, the x-axis points away from the Sun, and the z-axis is directed toward Galactic north.
With respect to the original version of the JF12 model, this improvement adopts a solenoidal transition for the disk field and parabolic field lines at $z=0$, instead of sharp kinks.

In order to compare our results to those from \cite{Chuzhoy:2008zy}, we also considered the model of a uniform Galactic magnetic field considered in that paper (Figure~\ref{fig:uniform}):
\begin{equation}
\label{eq:uniform}
\mathbf{B}_{\rm G}(z) =
  \begin{dcases}
B_{\rm homo} \mathbf{\hat{y}}
  & \mathrm{if}\, \, \,  
|z| \leq R,
 \\
0
  & \mathrm{else}\, \, \, .
 \end{dcases}
\end{equation}
Here $R$ is the size of the Galactic magnetic field. In the following, we also consider a finite version of this model for which the field is different from zero only for values of the $x$ and $y$ coordinates smaller than a distance $R'$:
\begin{equation}
\label{eq:uniform_finite}
\mathbf{B}_{\rm G}(z) =
  \begin{dcases}
B_{\rm homo} \mathbf{\hat{y}}
  & \mathrm{if}\, \, \,  
|z| \leq R,~|x|,|y| \leq R',
 \\
0
  & \mathrm{else}\, \, \, .
 \end{dcases}
\end{equation}
We underline that the models in Eqs.~\eqref{eq:uniform} and~\eqref{eq:uniform_finite} are  non-physical as they are respectively infinite in extent and not divergence-free.

\subsection{Gyromagnetic radius of CHAMPs}

The gyroradius of a CHAMP with mass $m_{\rm X}$ and charge $q_{\rm X}$, moving in a uniform magnetic field with velocity $v_{\rm X}$ can be expressed as \cite{Chuzhoy:2008zy}
\begin{equation}
\label{eq:gyro}
    R_{\rm g} \sim 10^{-9}\mathrm{pc} \left(\frac{m_{\rm X}}{m_{\rm p}} \right) \left(\frac{e}{q_{\rm X}} \right) \left(\frac{v_{\rm X}}{10^{-3}~c} \right) \left(\frac{B_{\rm G}}{1~\mu\mathrm{G}} \right)^{-1},
\end{equation}
It is claimed in \cite{Chuzhoy:2008zy} that CHAMPs in the Galactic halo cannot enter the Galactic disk unless $R_{\rm g}$ is larger than the typical height of the disk. Assuming a height of the disk around $100~\mathrm{pc}$, a velocity $v_{\rm X} = 10^{-3}~c$, and a magnetic field amplitude $B_{\rm G}=1~\mu\mathrm{G}$, they obtain that CHAMPs with masses smaller than $10^8~(q_{\rm X}/e)~\mathrm{TeV}$ cannot enter the Milky Way disk from the diffuse halo. This would nullify any attempt for direct detection of CHAMPs in terrestrial experiments.

In \cite{Chuzhoy:2008zy}, the authors implicitly assume a homogeneous Galactic magnetic field infinitely extended in the  x and y directions, as in Eq.~\eqref{eq:uniform}.
One problem in their argument is that, unlike the realistic situation with a finite size Galactic magnetic field, it is impossible to choose an initial connected set of particles that are all outside the magnetic region.
In the next section, we show that considering models with compact magnetic regions, as the model in Eq.~\eqref{eq:toroidal}, the JF12 model or even the model in Eq.~\eqref{eq:uniform_finite} with finite size of the magnetic region, CHAMPs with mass $\lesssim 10^8~(q_{\rm X}/e)~\mathrm{TeV}$ can actually reach the Earth, as there are always allowed trajectories from the diffuse halo to the Earth position in the Galactic disk.

\section{Simulation of the trajectories with CRPropa}
\label{sec:simulation}

In this section, we demonstrate that the results of \cite{Chuzhoy:2008zy} depend in an essential way on the adopted nonphysical model of the Galactic magnetic fields by simulating the motion of a gas of heavy charged particles in the models of Galactic magnetic fields considered in Section~\ref{sec:models}. In the simulations we neglect any gravitational effect, as also done in \cite{Chuzhoy:2008zy}, and therefore the particles move at constant speed. 

We adopt the simulation framework of CRPropa \cite{AlvesBatista:2022vem}, implementing in the code the possibility of a customized particle with a given mass and electric charge. CRPropa assumes particles moving at the light speed. Therefore, we have also modified the program in order to simulate also non-relativistic particles.

\subsection{Back-tracking uniform distributions at the Earth surface}

\begin{figure}
    \centering
    \begin{subfigure}[b]{0.49\textwidth}
         \centering
         \includegraphics[width=\textwidth]{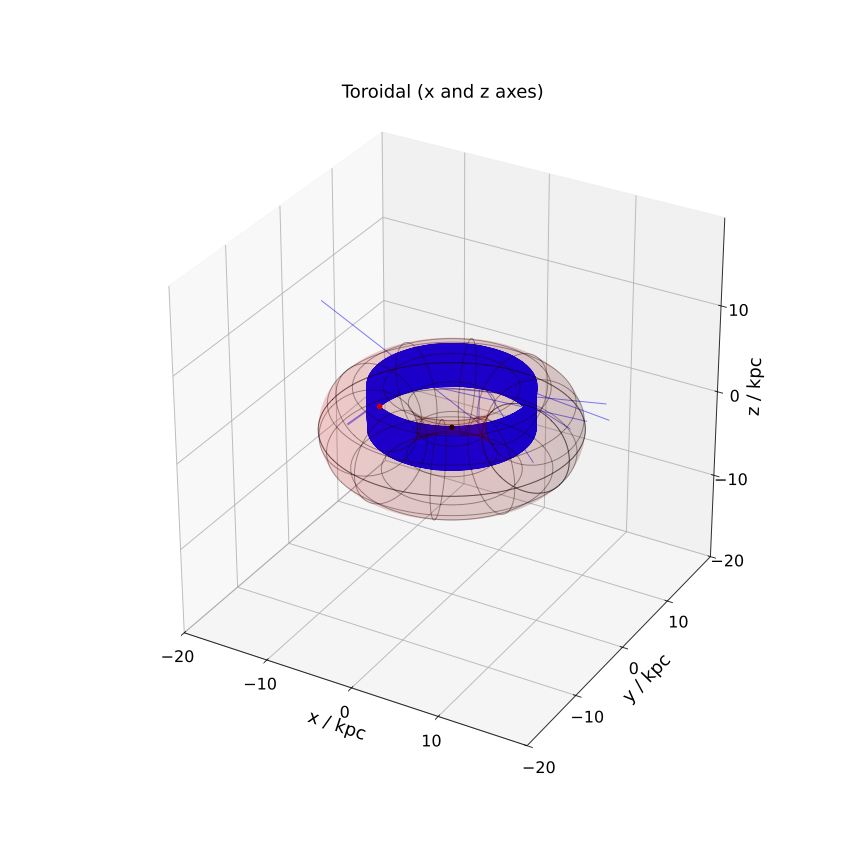}
         \caption{}
         \label{fig:particle_tracking_sol}
     \end{subfigure}
     \hfill
     \begin{subfigure}[b]{0.49\textwidth}
         \centering
         \includegraphics[width=\textwidth]{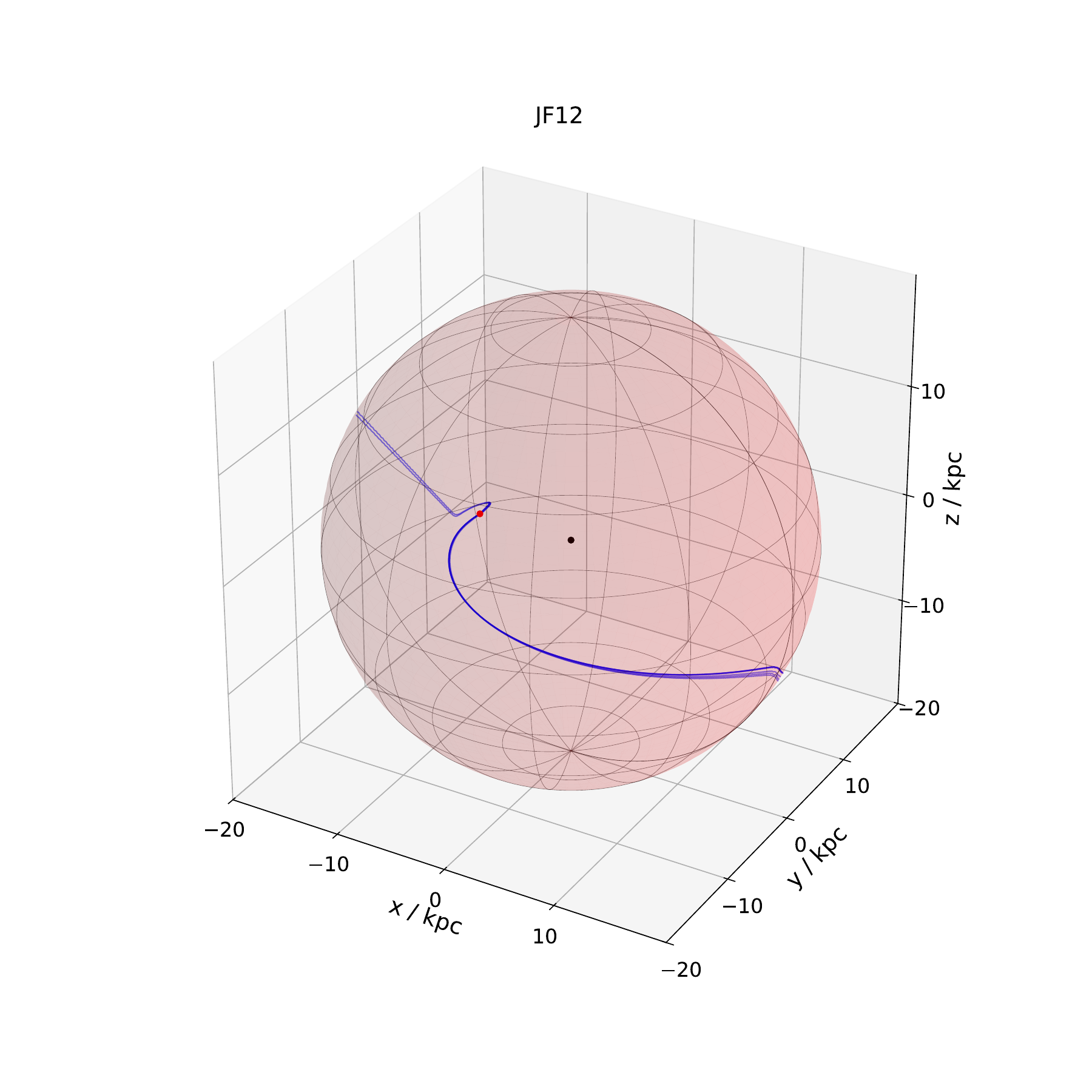}
         \caption{}
         \label{fig:particle_tracking_jf12}
     \end{subfigure}
     \hfill
     \begin{subfigure}[b]{0.49\textwidth}
         \centering
         \includegraphics[width=\textwidth]{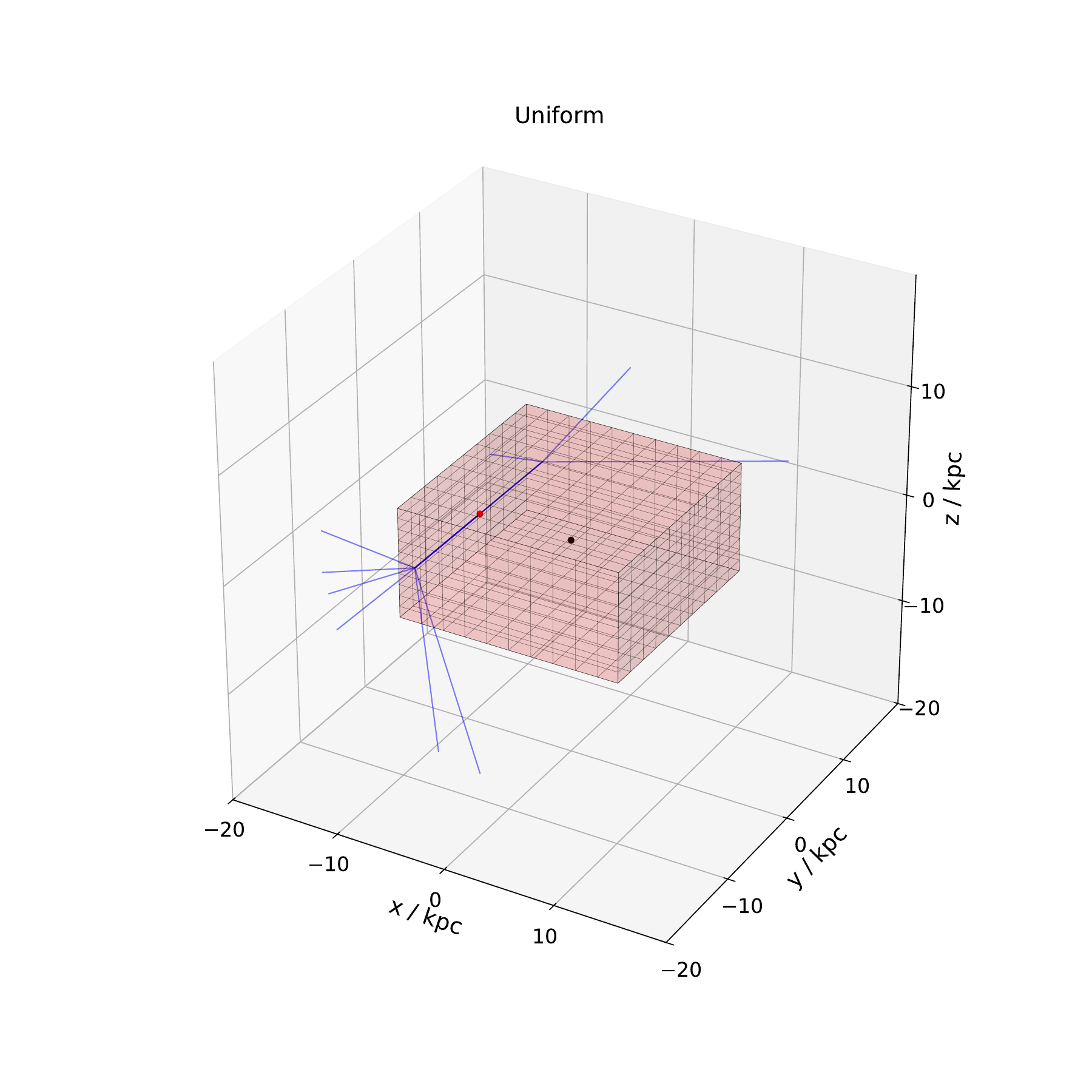}
         \caption{}
         \label{fig:particle_tracking_homo}
     \end{subfigure}
     \caption{Trajectories of CHAMPs with mass $10^7$~TeV and velocity $10^{-3} c$ that reach the Earth with an isotropic distribution. Here we show the simulation of 9 particles for three different models of the Galactic magnetic field:
     (a) toroidal solenoid in Eq.~\eqref{eq:toroidal} with $B_{\rm toro} = 2 \times 10^{-6}~\mathrm{G}$, $R_1 = 5~\mathrm{kpc}$, $R_2 = 8.5~\mathrm{kpc}$, (b) JF12 model \cite{Jansson_2012, Kleimann_2019}, (c) finite uniform model in \eqref{eq:uniform_finite} with $B_{\rm homo} = 2 \times 10^{-6}~\mathrm{G}$, $R = 5~\mathrm{kpc}$, $R' = 10~\mathrm{kpc}$.}
     \label{fig:particle_tracking}
\end{figure}

In Figure~\ref{fig:particle_tracking} we show the results of the simulations obtained by back-tracking a uniform distribution of cosmic rays as seen at the observer, which we identify with the Earth position, to a sphere of $20~\mathrm{kpc}$ around the Galactic center. We display the simulated trajectories of 9 particles for three different models of the Galactic magnetic fields: the toroidal model in Eq.~\eqref{eq:toroidal}, the JF12 model \cite{Jansson_2012, Kleimann_2019}, and the finite uniform model in Eq.~\eqref{eq:uniform_finite}. The particles have mass $m_{\rm X} = 10^7~\mathrm{TeV}$, charge $q_{\rm X} = e$, and velocity $v = 10^{-3} c$, which corresponds to the virial velocity of the Milky Way. In the plots, we also show the position of the Galactic center (black dot, $z=0$, $r=0$) and of the observer (red dot, $z=0$, $r=8.5~\mathrm{kpc}$).

For the toroidal model, we assume in Eq.~\eqref{eq:toroidal} $B_{\rm toro} = 2 \times 10^{-6}~\mathrm{G}$, $R_1 = 5~\mathrm{kpc}$, and $R_2 = 8.5~\mathrm{kpc}$. The results are shown in Figure~\ref{fig:particle_tracking_sol}. In the plot, we also show the boundaries of the magnetic region (red torus). As one can notice from the figure, in this model the particles reach the observer only from above the Galactic plane. This is due to the peculiar structure of the magnetic field, which in this model we take coherently counterclockwise, and on the fact that in the simulation we assume only particles with positive charge. Vice versa, particles with negative charge would reach the observer only from below the Galactic plane. Therefore, once a globally neutral distribution of particles and antiparticles is considered, the symmetry is restored.

In the case of the JF12 model (Figure~\ref{fig:particle_tracking_jf12}), using the same $m_X, q_X$ and $v$, the CHAMPs reaching Earth isotropically originate from two small regions on the $20~\mathrm{kpc}$ sphere, one in each of the northern and southern hemispheres.

Finally, in the case of the uniform magnetic field model (Figure~\ref{fig:particle_tracking_homo}), particles reach the observer only if they enter the Galactic magnetic region from a projected distance to the observer on the $xz$-plane equal or smaller than the gyroradius of the particles. 
Consequently, in the case of a magnetic region of infinite extent in the $xy$-plane, as assumed in \cite{Chuzhoy:2008zy}, only particles that are already inside the magnetic region can reach the observer, in agreement with the result of \cite{Chuzhoy:2008zy}.
For the simulation, we assumed $B_{\rm homo} = 2 \times 10^{-6}~\mathrm{G}$, $R = 5~\mathrm{kpc}$ and $R' = 10~\mathrm{kpc}$.

\begin{figure}
    \centering
    \begin{subfigure}[b]{0.49\textwidth}
         \centering
         \includegraphics[width=\textwidth]{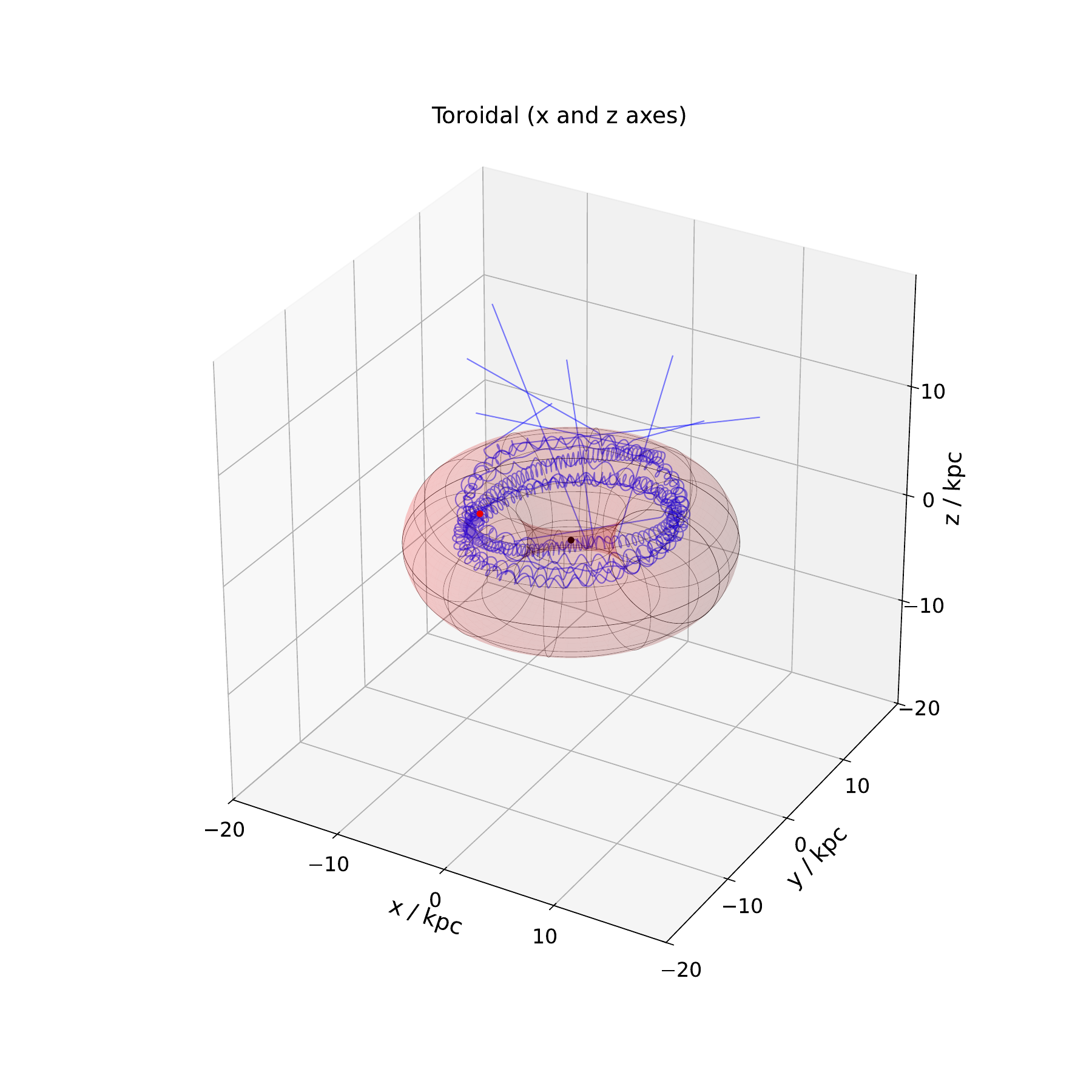}
         \caption{}
         \label{fig:particle_tracking_sol_1}
     \end{subfigure}
     \hfill
     \begin{subfigure}[b]{0.49\textwidth}
         \centering
         \includegraphics[width=\textwidth]{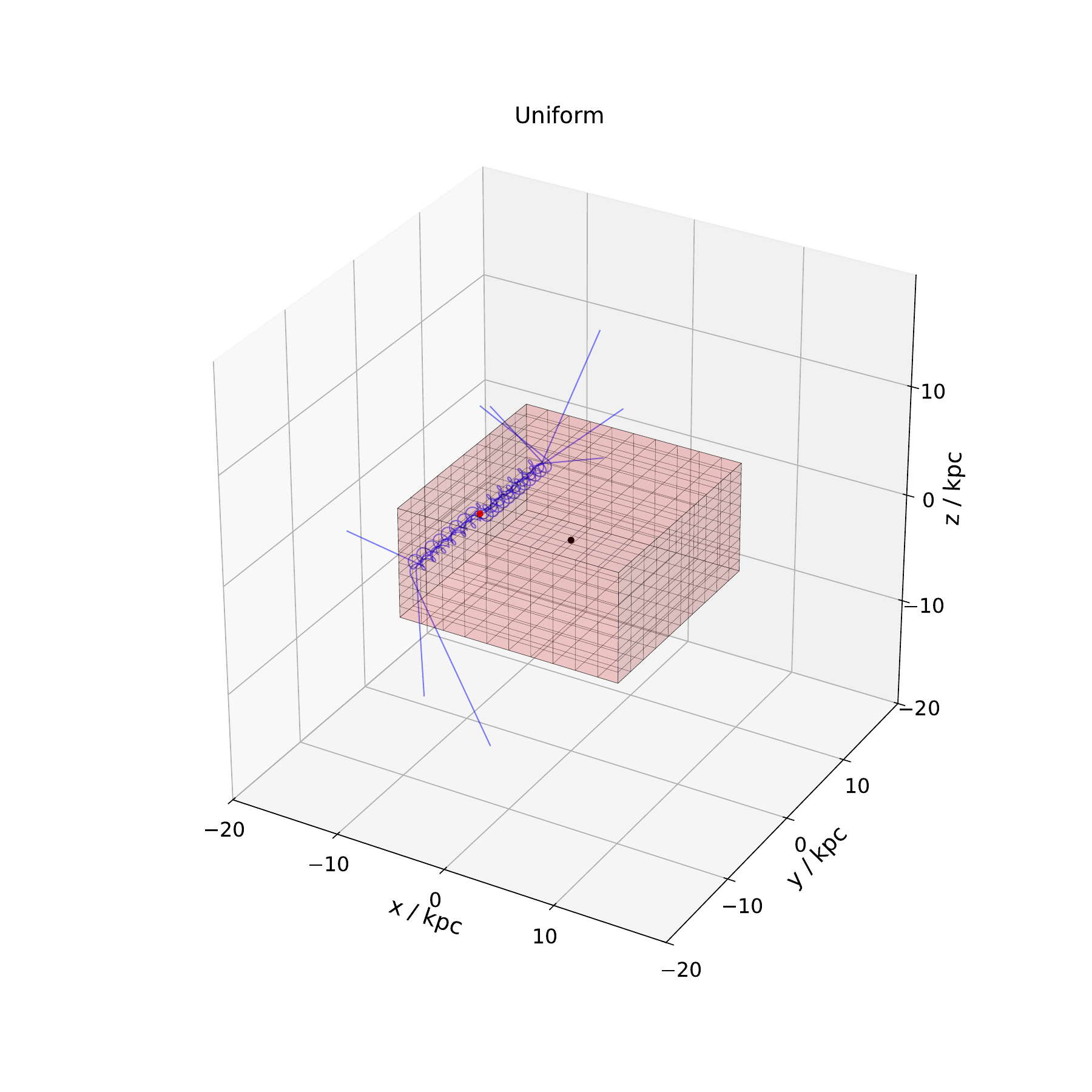}
         \caption{}
         \label{fig:particle_tracking_homo_1}
     \end{subfigure}
     \caption{Trajectories of heavy charged particles with mass $10^7$~TeV and velocity $10^{-1} c$ that reach the Earth with an isotropic distribution. Here we show the simulation of 9 particles for three different models of the Galactic magnetic field:
     (a) toroidal solenoid in Eq.~\eqref{eq:toroidal} with $B_{\rm toro} = 2 \times 10^{-6}~\mathrm{G}$, $R_1 = 5~\mathrm{kpc}$, $R_2 = 8.5~\mathrm{kpc}$, (b) finite uniform model in \eqref{eq:uniform_finite} with $B_{\rm homo} = 2 \times 10^{-6}~\mathrm{G}$, $R = 5~\mathrm{kpc}$, $R' = 10~\mathrm{kpc}$.}
     \label{fig:particle_tracking_1}
\end{figure}

For the realistic CHAMP parameters adopted in these simulations, the expression for the gyroradius (in the approximation of constant magnetic field) in Eq.~\eqref{eq:gyro} gives $R_{\rm g} \sim 5~\mathrm{pc}$ for incident velocity perpendicular to the Galactic magnetic field and, hence, the gyroradius is too small to be visible in Figure~\ref{fig:particle_tracking}.
Therefore, to confirm the validity the simulation results, we simulate the motion of particles with the same mass and charge, but velocity $v = 10^{-1} c$, back-tracking a homogeneous distribution of particles at the Earth in the case of the toroidal and finite uniform field models. We show the results in Figure~\ref{fig:particle_tracking_1}.
From the expression in Eq.~\eqref{eq:gyro}, for a uniform magnetic field of $B_{\rm homo} =  2 \times 10^{-6}~\mathrm{G}$ such particles have a gyroradius $R_{\rm g} \sim 500~\mathrm{pc}$. 
In this case, it is possible to observe in the plot the size of the gyromagnetic radius of the particles, which is consistent with the result in Eq.~\eqref{eq:gyro}.
According to the argument of \cite{Chuzhoy:2008zy}, for a disk height of $5~\mathrm{kpc}$, as in the simulation, these particles should not be able to reach the Earth from outside the Galactic disk. As we saw above, in the unrealistic case of a planar field of infinite extent along the direction it is pointing, this is true.  However when this is truncated to have finite extent (an unacceptable not-divergence-free example) CHAMPs from outside the disk do enter. The results demonstrate yet again that the claim of~\cite{Chuzhoy:2008zy}, that CHAMPs in a certain mass range cannot enter the magnetized disk of the Galaxy, is not valid for physically acceptable field models.

\subsection{Angular distribution of CHAMPs at the Earth}

While the simulations of the previous section were obtained by back-tracking the particles, in this section we also simulate the evolution of a homogeneous and isotropic ingoing distribution of CHAMPs with a mass of $m_{\rm X} = 10^7~\mathrm{TeV}$ and charge $q_{\rm X} = e$ on a spherical surface of radius $20~\mathrm{kpc}$ around the Galactic center. 
Considering that from an isotropic propagation in the extragalactic Universe, the particle directions distributed on a spherical surface follow Lambert's distribution\footnote{This can be understood by considering that vertical incident angles are more frequent due to the larger visible size of the area of the surface element than for nearer-horizontal angles.}, we assume for the particles on the sphere a homogeneous distribution with arbitrary initial direction following Lambert's distribution. In Appendix~\ref{sec:number_estimate}, we describe the distribution in details.
During the simulation we drop the particles that miss the Earth and exit from a spherical surface of radius $25~\mathrm{kpc}$ centered in the Galactic center. Moreover, we adopt the condition upon detection that the particle must be at some time within a distance from the observer position at the Earth smaller than $300~\mathrm{pc}$. The radius of the sphere has been chosen to get a significant statistics of particles.

\begin{figure}
    \centering
    \includegraphics[width=\linewidth]{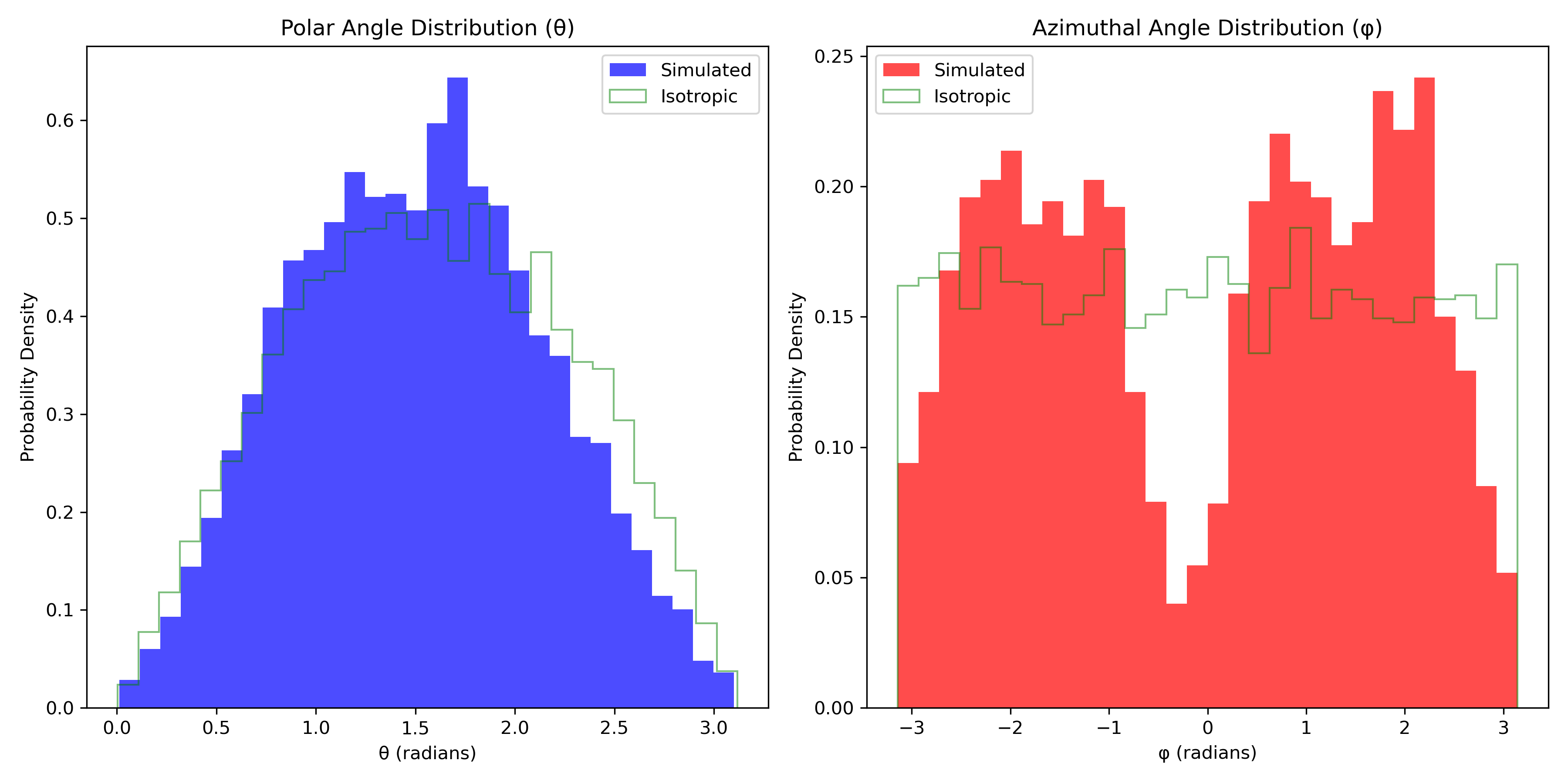}
    \caption{Angular distribution of the incoming particles on Earth, assuming a Lambert distribution on the surface of a sphere of radius $20~\mathrm{kpc}$ and propagating through the fully divergence-free version of the JF12 magnetic field model of~\cite{Kleimann_2019} with an initial sample of $10^8$ particles. The light blue line shows the same, for an isotropic distribution with the same number of events. A particle mass of $m_{\rm X} = 10^7$~TeV and velocity $v = 10^{-3} c$ are assumed. The results of the simulation clearly show the presence of two different populations of particles. See the main text for a discussion of the anisotropies in the simulated distribution.
    }
    \label{fig:angular_JF12}
\end{figure}
In Figure~\ref{fig:angular_JF12} we show the angular distribution of the simulated particles in the case of the JF12 model. We consider an initial number of particles ($10^8$) moving at velocity $v = 10^{-3} c$. 
At the end of the simulation 6460 particles have reached the observer, while all the others have been dropped because they exited the spherical surface of radius $25~\mathrm{kpc}$. This result agrees at the order-of-magnitude level, with the fraction of particles that would have reached the observer from the spherical surface in the absence of any magnetic field, that is $f \sim 10^{-4}$ (see Appendix~\ref{sec:number_estimate} for the details of the computation).
In the figure we also show a randomly generated isotropic distribution at the Earth with the same number of events that reach the observer.
In this case, the gyroradius of the particles ($\sim 5~\mathrm{pc}$) is much smaller than the size of the observer and therefore the angular distribution is only sensitive to the particle direction perpendicular to the magnetic field.
As already noticed in Figure~\ref{fig:particle_tracking_jf12}, from the distribution of the azimuthal angle the existence of two different populations reaching the Earth from two different directions is evident. This corresponds to the two visibly different populations in the right plot of the figure. The width of the distribution is related to the size of the sphere surface of the observer. The same effect can be observed for the polar angle distribution where the distribution prefers smaller polar angle with respect to the isotropic distribution.

\begin{figure}
    \centering
    \includegraphics[width=\linewidth]{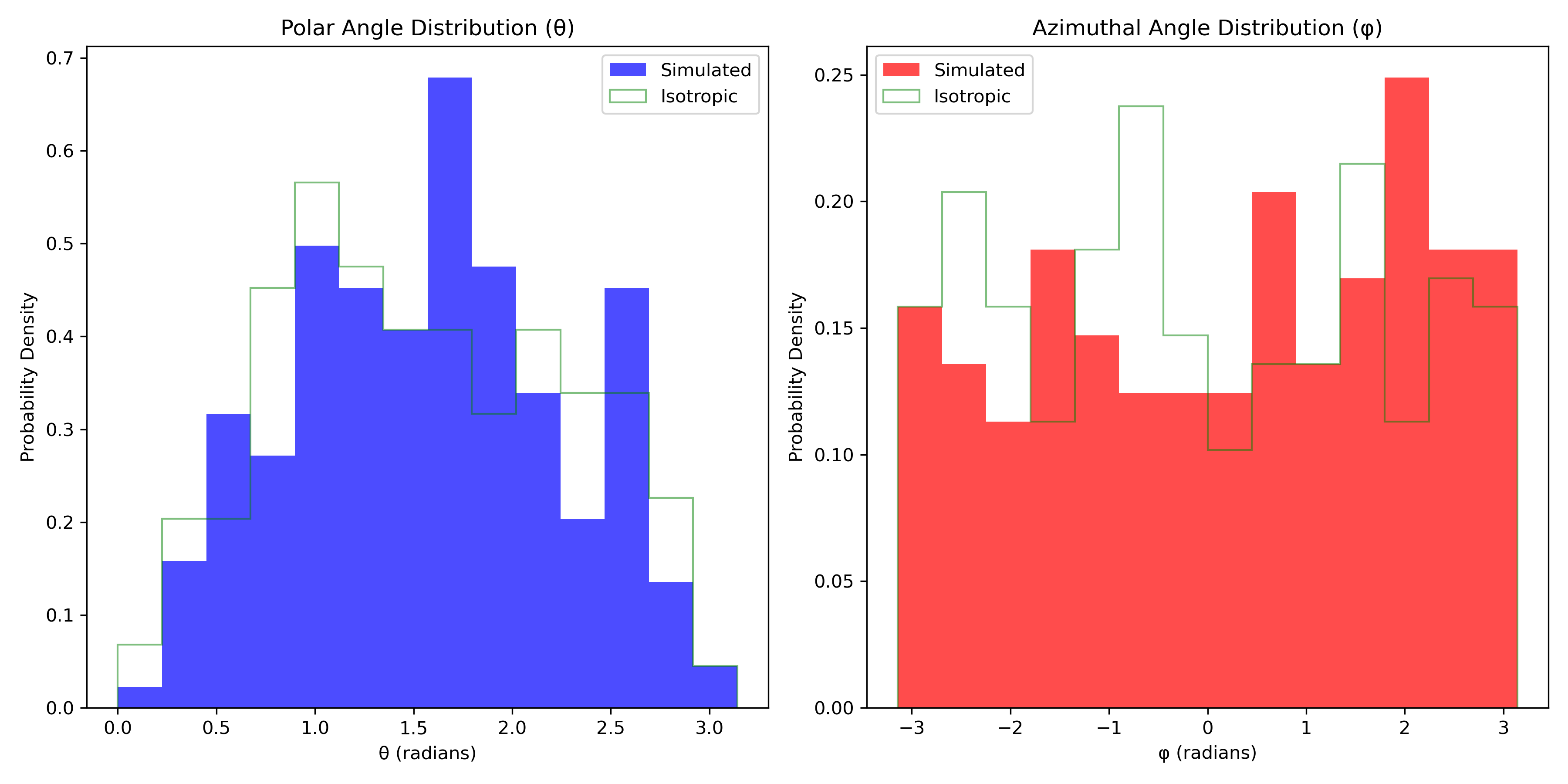}
    \caption{Angular distribution of the incoming particles on Earth, assuming Lambert's distribution on the surface of a sphere with radius $20~\mathrm{kpc}$ and the toroidal solenoid magnetic field model for the Milky Way. Here we consider an initial number of particles of $10^6$. We also show the result for an isotropic distribution with the same number of events. We assume for the particles a mass $m_{\rm X} = 10^7$~TeV and a velocity $v = 10^{-1}~c$. The results of the simulation are compatible with an isotropic distribution, see the main text for more details.}
    \label{fig:angular_solenoid}
\end{figure}
In Figure~\ref{fig:angular_solenoid}, we show the angular distribution of the particles in the toroidal model. As in Figure~\ref{fig:particle_tracking_1}, we take a particle velocity $v = 10^{-1}$, with an initial sample of $10^6$ particles.\footnote{A smaller number of particles was simulated here because the toroidal model requires more computing time due to the peculiar trajectories shown in Figure~\ref{fig:particle_tracking_sol_1}.}
In this case, the gyroradius of the particles ($\sim 500~\mathrm{pc}$) is larger than the radius of the observer surface. Consequently, the simulation is sensitive not only to the particle motion along the magnetic field lines but also to their rotational motion.
Among the initial particles, only 197 reached the observer by the end of the simulation. 
As in the previous case, the result is consistent at the order-of-magnitude level with the fraction of particles expected to arrive at the observer from a spherical surface in the absence of magnetic fields, as estimated in Appendix~\ref{sec:number_estimate}.
We also display an isotropic distribution with the same number of events for comparison. 
The angular distribution of the simulated trajectories is consistent with isotropy. This behavior arises because, when the gyroradius exceeds the observer radius, the rotational motion of the particles effectively isotropizes their trajectories. We therefore expect a similar effect for the parameters used in the simulation of Figure~\ref{fig:angular_JF12} if the observer radius is smaller than the particle gyroradius.

\subsection{Discussion of the results}

The different simulations we have presented show how particle trajectories that reach the Earth from outside the Galactic disk are allowed for CHAMPs with $m_{\rm X}\lesssim 10^8~\mathrm{TeV}$, as long as the magnetic field region is physical, e.g. does not have the infinite planar geometry used in \cite{Chuzhoy:2008zy}.
In particular, the simulations demonstrate how, while we can reproduce the results of \cite{Chuzhoy:2008zy} for the model adopted by the authors, their hypothesis is falsified once we consider physically reasonable models for the Galactic magnetic field. 
This result can be intuitively understood as the fact that in the presence of magnetic fields, sources in some directions can be deflected without reaching the observer, while other sources that would not illuminate the observer without deflection contributes to the isotropy of the distribution.
Hence, the argument of \cite{Chuzhoy:2008zy} in favor of a CHAMP dark matter window holds only for unphysical models of the Galactic magnetic field, and is not generic.

We also observed that the expected number of particles reaching the observer at the end of the simulation is compatible at the order of magnitude level with the particle number expected in the absence of any magnetic field, showing that the statistic of the particles cannot be significantly altered by the fields.

\section{Conclusion}
\label{sec:conclusion}

In this work, we have reconsidered the possibility of a window for CHAMP dark matter within the mass range $100 (q_{\rm X}/e)^2~\mathrm{TeV} \leq m_{\rm X} \leq 10^8 (q_{\rm X}/e)~\mathrm{TeV}$, which was argued in \cite{Chuzhoy:2008zy} to be viable. 
We simulated the motion of a gas of CHAMPs moving at constant velocity in different models of the Galactic magnetic field, obtaining results falsifying the argument of \cite{Chuzhoy:2008zy}.
We demonstrated that the results of \cite{Chuzhoy:2008zy} rely on their physically unacceptable model of an infinitely-extended magnetic field, and do not obtain with divergence-free, finite-size GMF models of the Galactic magnetic field.

\acknowledgments
The research of GRF has been supported by NSF grants PHY-2013199 and PHY-2413153.
D.P. was partially supported by the National Science Centre, Poland, under research grant no. 2020/38/E/ST2/00243.

\appendix

\section{Estimate of the number of particle reaching an observer from an incoming Lambert distribution}
\label{sec:number_estimate}

In this appendix, we first review the Lambert distribution and then estimate the number of particles reaching a spherical observer with radius $r_{\rm obs}$ from a spherical distribution of incoming particles with velocity direction on any surface element distributed following the Lambert distribution and radius $r_{\rm ext} > r_{\rm obs}$.

The Lambert distribution was first proposed in optics as the luminous intensity distribution from an ideal diffusely reflecting surface, which follows a cosine law of the angle between the observer's line of sight and the normal surface. In our case, we can express the Lambert distribution of particles emitted from a surface element in the ingoing hemisphere as
\begin{equation}
    P(\theta, \phi) d\Omega = \frac{\cos \theta}{\pi} d\Omega ,
\end{equation}
where $\theta \in (0,\pi/2)$ is the angle between the observer's line of sight and the normal surface, $\phi \in (0, 2\pi)$ is the azimuthal coordinate, and $d\Omega$ is the solid angle element. The $\pi$ factor at the denominator insures the normalization of the distribution.

We now estimate the number of particles that we expect to reach an inner sphere with radius $r_{\rm obs}$ from an ingoing distribution of particles following the Lambert distribution from a spherical surface with radius $r_{\rm ext} > r_{\rm obs}$, in the absence of any magnetic field.
To estimate the fraction of particles, $f$, we compute the effective solid angle that the target subtends as seen by the spherical surface, $\Omega_{\rm eff}$:
\begin{equation}
    f = \int_{\Omega_{\rm eff}} \frac{\cos \theta}{\pi} d\Omega \sim \frac{\Omega_{\rm eff}}{2 \pi} ,
\end{equation}
where the second equality holds under the assumption of small $\Omega_{\rm eff}$.
In the small-angle approximation, we estimate the effective solid angle as
\begin{equation}
    \Omega_{\rm eff} \sim \pi \left( \frac{r_{\rm obs}}{\langle r \rangle} \right)^2 ,
\end{equation}
with $\langle r \rangle \sim r_{\rm ext}$ being the average distance of the inner sphere from the outer spherical surface. For the parameters adopted in Section~\ref{sec:simulation}, $r_{\rm obs} = 300~\mathrm{pc}$ and $r_{\rm ext} = 20~\mathrm{kpc}$, we estimate a fraction of particles hitting the observer sphere over the total of emitted particles to be of order $f \sim 10^{-4}$.

\bibliographystyle{unsrt}
\bibliography{biblio}

\end{document}